\documentclass[twocolumn,showpacs,prc]{revtex4}
\usepackage{graphicx}
\usepackage{dcolumn}
\usepackage{bm}
\begin{document} 

\title{Dynamics of zero-point energy and two-slit phenomena for photons}
\author{A. Widom and J. Swain}
\affiliation{Physics Department, Northeastern University, Boston MA USA}
\author{Y. N. Srivastava}
\affiliation{Department of Physics and Geology, University of Perugia, Perugia, Italy}
\author{M. Blasone and G. Vitiello}
\affiliation{Dipartimento di Fisica, Universit\`a di Salerno, 84100 Salerno, Italy} 
\affiliation{Gruppo Collegato INFN Salerno, Sezione di Napoli, Napoli, Italy}

\begin{abstract}
An earlier forward and backward in time formalism developed by us to discuss non-relativistic electron
diffraction is generalized to the relativistic case and here applied to photons. We show how naturally the
zero-point energy emerges in the Planck black-body spectrum once symmetric in time motion - inherent in the
Maxwell equations - is invoked for  photons. Then, a detailed study is made of two-slit experiments for photons and 
some novel phenomena, amenable to experiments, are proposed, that arise due to the spin of the photon.
\end{abstract}

\pacs{82.47.Jk, 82.47.Uv, 84.60.Jt}
\maketitle

\section{Introduction \label{intro}}
In his study of the Brownian motion of a quantum oscillator, Schwinger introduced the notion of coordinates
moving forward in time, $x_+(t)$, and coordinates $x_-(t)$  moving backward in time \cite{Schwinger}.
By using such a doubling of the degrees of freedom, Schwinger developed in full generality a mathematically
complete formalism for dealing with quantum Brownian motion. The starting point in his analysis is that a
quantum object may be viewed as splitting the single coordinate, say $x(t)$,  into two coordinates $x_+
(t)$ (going forward in time) and $x_- (t)$  (going backward in time). From the Schwinger quantum operator
action principle it can be derived that the classical limit is obtained when both motions coincide
$x(t)=x_+ (t)=x_- (t)$. The impact of Schwinger's notion of forward \& backward in time coordinates on
subsequent studies in stochastic mechanics, many-body physics, quantum dissipation and thermal quantum
field theory (QFT) in general, has been enormous. Such a concept has been also used in order to illustrate
the non-relativistic electron beam two-slit diffraction experiments in  \cite{Blasone:1998,Siva3:2003}. 
The interference patterns were there computed with or without dissipation (described by a thermal bath).
A dissipative interference phase, due to the inherent non-commutative geometry, closely
analogous to the Aharanov-Bohm magnetic field induced phase, was also found.

Proceeding further, 
in \cite{Siva1:2004},  
using Maxwell's equations  the photon {\em Zitterbewegung} motion along helical paths was
explored and the resulting non-commutative
geometry of photon position and, the distance between two photons in a
polarized beam of a given helicity was shown to have a discrete spectrum
that should become manifest in measurements of two photon coincidence counts. An experiment was proposed and its
feasibility examined in \cite{Siva2:2004}.

In the present paper, we extend Schwinger's formalism of forward \& backward  in time motions, used
in \cite{Blasone:1998,Siva3:2003}, to the relativistic case of the photon field.

Our discussion will proceed actually in two parts. In the first part we will focus our analysis on the
contribution to the zero-point energy of the forward \& backward in time motions of field modes. In the
second part, we will consider more specifically the two-slit photon experiments and propose a novel set of experiments.

In Sec.~\ref{Maxwell}, we review the Maxwell equations to emphasize
that Maxwell's theoretical construction is inherently time symmetric. This is illustrated by showing that the
zero-point energy in the Planck black-body spectrum finds its natural explanation once the forward \&
backward  time-symmetry is enforced \cite{Widom:2015}. Since the photon is its own antiparticle, the notion of 
time-symmetry is often obscured. We illustrate it in Sec.~\ref{pion} by considering the case of a spin zero, 
charged (boson) field for which the two motions are distinct. Extension to any integer and half-integer spin is also considered.

While the emphasis in \cite{Siva1:2004,Siva2:2004} was upon the non-commutative photon field
coordinates and on methods for its revelation through two-photon processes, here in Sec.~\ref{photo},
turning to the second part of our discussion, we shall be focusing on the behavior of a single photon for
two-slit arrangements to shed light on forward and backward in time propagation.

In Sec.~\ref{spinless}, we first discuss photons as scalar (spin zero) fields and deduce for it the well-known
diffraction pattern as in classical optics.
 However, once non-commuting spins are introduced, the quantum and
classical theories need not be equivalent. For example, the spin precession needs to be considered \cite{Siva1:2004}. This is studied in Sec.
\ref{spin1} and indeed a novel 
constraint - not present for scalar fields - is found when the slit width $w < \lambda$, the wave-length of the radiation.
Our proposal concerns
experiments in such a limit which, as far as we know, has not been investigated.
Further experimental issues are discussed in
Sec.~\ref{past}.

Conclusions are presented in Sec.~\ref{Con} and some details of the formalism are given in the Appendices.

\section{Maxwell Forward \& Backward in Time Motion and Zero-point Energy in Planck Black-body Radiation
\label{Maxwell}}

The present Section and the 
next are devoted to 
the zero-point energy generated by 
the forward \& backward in time motions of field modes. 
For this purpose, we first review the 
time symmetric character of Maxwell equations. We show that 
the zero-point energy in the Planck black body spectrum is 
due to the symmetric forward and backward in time motion of photons.
Otherwise said, due to the symmetric distribution in the photon frequency $\omega \leftrightarrow - \omega$.
Our discussion is mostly based on derivations presented in \cite{Widom:2015}. 

For pure radiation, i.e., in a part of space-time that is devoid of charges and currents,  
 the Maxwell field equations read
\begin{eqnarray}
\label{3}
{\bf \nabla}\cdot {\bf E} &=& 0;\\ {\bf \nabla}\cdot {\bf B} &=& 0;\\ {\bf \nabla} \times {\bf E} &=& -
\frac{1}{c} \,\frac{\partial {\bf B}}{\partial t} ;\\ {\bf \nabla} \times {\bf B} &=&  \frac{1}{c}
\,\frac{\partial {\bf E}}{\partial t} .
\end{eqnarray}
As the (positive definite) Maxwell EM energy density \cite{Maxwell:1878} is proportional to ($\bf{E}^2 +
\bf{B}^2$), it is natural to associate it with (${\bf F}\cdot {\bf F}^\dagger$), where the complex vector
fields are chosen as
\begin{eqnarray}
{\bf F}={\bf E} + i{\bf B};\ \ \quad  {\bf F}^\dagger={\bf E} - i{\bf B};
\\
{\bf \nabla}\cdot {\bf F}=0; ~~~ ~~ \quad  {\bf \nabla}\cdot {\bf F^\dagger} = 0.
\label{pf2}
\end{eqnarray}

One also sees that
\begin{equation}
{\bf F}\cdot {\bf F} = |{\bf E}|^2-|{\bf B}|^2+2i{\bf E\cdot B},
\label{pf3}
\end{equation}
which determines the Lorentz scalar (\begin{math} |{\bf E}|^2-|{\bf B}|^2 \end{math}) and Lorentz
pseudo-scalar (\begin{math} {\bf E\cdot B} \end{math}).

Using  Maxwell's equations, it is easy to show that ${\bf F, F^\dagger}$ obey the Schr${\ddot o}$dinger
equation, along with the transversality condition:
\begin{eqnarray}
\label{5}
i \hbar \frac{\partial F_j}{\partial t} = (\hbar c) \epsilon_{jkl} \partial_k F_l \equiv\ H^{(+)}_{jk}
F_k;\\
i\hbar \frac{\partial F^\dagger_j}{\partial t} = -(\hbar c) \epsilon_{jkl} \partial_k F^\dagger_l \equiv\
H^{(-)}_{jk} F^\dagger_k; \label{5b}\\
\partial_j F_j = 0;       \qquad      \partial_j F^\dagger_j = 0.
\end{eqnarray}
Define, the momentum operator $p_j = -i \hbar \partial_j$ and a spin-one operator ${\bf S}$ with matrix
elements $(S_j)_{kl} = -i \epsilon_{jkl}$, with ${\bf S}^2 = s(s+1) = 2$, so that we may rewrite
Eqs.(\ref{5}) and (\ref{5b}) as matrix equations
\begin{eqnarray}
\label{6a}
i \hbar \frac{\partial{\bf F}}{\partial t} &= &c \ ({\bf p \cdot S}) {\bf F} = H^{(+)} {\bf F}; \\
i \hbar \frac{\partial{\bf F^\dagger}}{\partial t} &=& - c \ ({\bf p \cdot S}) {\bf F^\dagger} = H^{(-)}
{\bf F^\dagger} \label{6}.
\end{eqnarray}
Physically, in this ({\it Maxwell}) representation, ${\bf F}$ goes forward in time and ${\bf F^\dagger}$
goes backward in time.

Eqs.(\ref{6a}), (\ref{6}) may be written more compactly as a Schr${\ddot o}$dinger equation in a 6-component form by
putting
\begin{equation}
\label{7}
 \Psi = \left(\begin{array}{ccc}
{\bf F}\\ {\bf F}^\dagger
\end{array} \right),
\end{equation}
and
\begin{equation}
\label{8}
\beta = \left(\begin{array}{ccc}
1 &0\\ 0 &-1
\end{array} \right);    \qquad
\ {\bf \Gamma} = \beta\ {\bf S},
\end{equation}
so that
\begin{equation}
\label{9}
i \hbar \frac{\partial \Psi}{\partial t} = c\ ({\bf p \cdot \Gamma}) \Psi \equiv\ \mathcal{H} \Psi.
\end{equation}
It is clear that the eigenvalues $\pm 1$ of $\beta$ distinguish the forward versus backward in time
motions. Explicitly
\begin{eqnarray}
\label{10}
&&\hspace{-3mm}\Psi =  \frac{1 + \beta}{2} \, \Psi ~ + ~ \frac{1 - \beta}{2} \, \Psi \, \equiv\ \, \Psi_+ ~
+ ~  \Psi_-;\\ 
&&\hspace{-5mm}\Longrightarrow\ \mathcal{H} \Psi_+ \equiv\ H^{(+)} \Psi_+;   \quad \mathcal{H} \Psi_-
\equiv\ H^{(-)} \Psi_- ;
\\ &&\hspace{2mm}H^{(+)} = - H^{(-)} = c\ {\bf p \cdot S} \ .
\end{eqnarray}
Much of the above formalism  can be found in
\cite{Siva1:2004}.

 A symmetric treatment of forward and backward in time motions is part and
parcel of the Maxwell field theory. In the following, we shall show that once this
intrinsic time symmetry in the Maxwell equation is enforced, the zero-point energy in the Planck black body
thermal radiation follows.

Let us recall that Planck originally \cite{Planck:1901} discussed the mean number of photons
of frequency \begin{math} \omega \end{math} in the thermal vacuum
\begin{equation}
\bar{n}= \frac{1}{e^{\hbar \omega /k_BT}-1} .
\label{pd1}
\end{equation}
The mean thermal energy of an electromagnetic oscillator was thereby taken to be
\begin{equation}
\bar{E}(\omega )=\hbar \omega \bar{n}
= \frac{\hbar \omega }{e^{\hbar \omega /k_BT}-1} .
\label{pd2}
\end{equation}
Later \cite{Planck:1912} Planck arbitrarily added the zero-point energy:
\begin{equation}
E_T(\omega )=\hbar \omega \left(\bar{n}+\frac{1}{2}\right) = \frac{\hbar \omega }{2} \,
\coth \left(\frac{\hbar \omega }{2k_BT}\right).
\label{pd2a}
\end{equation}
Remarkably, Eq.(\ref{pd2a}) is symmetric in $\omega \leftrightarrow -\omega$. 
Of course, as well known, the zero-point energy is obtained by actually
solving the quantum mechanical  harmonic oscillator. 

 Einstein and Stern \cite{Einstein:1913} noted that the excess energy over
and above the equipartition value obeyed
\begin{equation}
\lim_{T\to \infty}
\left(\bar{E}(\omega )+\frac{\hbar \omega}{2}-k_BT\right) =0,
\label{pd2b}
\end{equation}
that might theoretically be regarded as slight evidence of a zero temperature energy of \begin{math}
\hbar \omega /2   \end{math}.

We now remark that
\begin{equation}
E_T(\omega )=\frac{1}{2}\left[\bar{E}(\omega )+\bar{E}(-\omega )\right].
\label{pd4}
\end{equation}
Eq.(\ref{pd4}) is indeed true in virtue of Eqs.(\ref{pd2}) and (\ref{pd2a}). Also, note
the zero-point  energy
\begin{equation}
E_0(\omega ) \equiv \lim_{T\to 0^+} E_T(\omega ) = \frac{\hbar |\omega |}{2} .
\label{pd5}
\end{equation}
The relevance of Eqs.(\ref{pd4}) and (\ref{pd5}) relies in the fact that they exhibit the contributions to
the zero-point  energy by the positive and negative frequency modes (forward and backward in time,
respectively). Of course, if one expresses this result in terms of the photon creation operator
\begin{math} a^\dagger \end{math} and destruction operator
\begin{math} a \end{math} with
\begin{math} \big[a,a^\dagger \big]=1 \end{math}, then the photon
number operator \begin{math} n=a^\dagger a  \end{math} enters into the Hamiltonian via the symmetrized
product
\begin{math} \big(a a^\dagger + a^\dagger a \big) \end{math} as
\begin{equation}
{\cal H}=\frac{\hbar |\omega |}{2}\left(a^\dagger a + a a^\dagger \right) =\hbar |\omega
|\left(n+\frac{1}{2}\right).
\label{pd6}
\end{equation}
Eq.(\ref{pd6}) leads directly to Eq.(\ref{pd5}).

In conclusion, the physical meaning of Eq.(\ref{pd4}) is that both, the positive frequency \begin{math}
\omega > 0  \end{math},  a particle moving forward in time, and the negative frequency
\begin{math} \omega < 0  \end{math},  an anti-particle moving
backward in time, contribute to the zero-point  energy. Since the photon is its own anti-particle, the
physical meaning of Eqs.(\ref{pd4}) and (\ref{pd5}) may be somewhat obscured. In order to make the particle
content in the zero-point  energy more evident, we consider in the following Section \ref{pion}, a case wherein the
particle and anti-particle are distinct.

\section{Charged fields  \label{pion}}

In this Section we discuss first 
spinless charged boson oscillator energies. Extension
to the non-zero spin boson and fermion field 
are taken up later in the second part of subsection  \ref{cpp}
(cf. Eq.(\ref{cpp12d})).

The energy of a spinless charged boson field  in a uniform magnetic field \begin{math} {\bf
B}=(0,0,B)=(0,0,|{\bf B}|) \end{math} is given by \cite{Landau:1980}
\begin{eqnarray}
\epsilon_\pm (n,p,B)=\pm c\sqrt{m^2c^2 + p^2  +(2n+1)|\hbar e{\bf B}|/c}\ ,
\label{pion1}
\end{eqnarray}
wherein the integer \begin{math} n = 0,1,2 \dots \ \end{math} is the label for the circular Landau
orbit, the momentum along the magnetic field axis is \begin{math} {\bf p}=(0,0,p) , \, p=\hbar k \end{math}
and \begin{math} \kappa =(mc/\hbar )\end{math} is the mass in inverse length units. Thus
\begin{eqnarray}
\omega(n,k,B)=c\sqrt{\kappa^2 + k^2  +(2n+1)|e{\bf B}|/\hbar c}\ .
\label{pion2}
\end{eqnarray}
The zero-point  charged boson oscillator energies per unit volume counting the particle and anti-particle
separately in virtue of the different charge \begin{math} \pm e \end{math} is determined by
\begin{eqnarray}
U_0(B)=2\times
 \frac{eB}{2\pi \hbar c} \sum_{n=0}^\infty
\int_{-\infty}^\infty \frac{dk}{2\pi}
\frac{\hbar \omega(n,k,B)}{2}.
\label{pion3}
\end{eqnarray}
This vacuum energy per unit volume in a magnetic field is clearly divergent so one must {\em regularize}
and {\em renormalize}. After doing both 
exercises. a finite vacuum boson energy per unit volume in a
magnetic field
\begin{math} U(B) \end{math} arises. We present in the Appendix A some of the Gamma function regularization
formalism and we briefly comment on the charge renormalization procedure.

The physical fields are defined so that the normal vacuum magnetic energy density is
\begin{math} |{\bf B}|^2/8\pi  \end{math}. This can be realized by a charge
renormalization subtraction in Eq.(\ref{gfr8}). Thus, for scalar boson fields the vacuum energy density is
obtained as
\begin{eqnarray}
U(B)&=& \frac{\hbar c}{16\pi^2}
\int_0^\infty \frac{ds}{s^3}\ e^{-\kappa ^2s} \times
\nonumber \\
&&\left[1-\frac{(eBs/\hbar c)}{\sinh(eBs/\hbar c)}-
\frac{(eBs/\hbar c)^2}{6}\right].
\label{cr1}
\end{eqnarray}
Eq.(\ref{cr1}) is both finite and exact for the sum of zero-point  oscillations of charged boson spin zero
systems. 
The vacuum boson magnetization is thereby
\begin{equation}
{\bf M}=- \frac{\partial U}{\partial {\bf B}} .
\label{cr2}
\end{equation}

To consider now what happens in an external electric field,  we recall that to go from a pure external
magnetic field to a pure external electric field one takes \begin{math} B^2 \to -E^2 \end{math}. This
allows us to obtain the boson pair production rate
\begin{math} \Gamma \end{math} per unit time per unit volume in an external
electric field. This may be computed from
\begin{equation}
\Gamma = - \frac{2}{\hbar }
{\Im m}\ U(B\to -iE).
\label{bpp1}
\end{equation}
Eqs.(\ref{cr1}) and (\ref{bpp1}) imply
\begin{eqnarray}
\Gamma &=& \frac{c}{8\pi ^3}\left(\frac{eE}{\hbar c}\right)^2 \times
\nonumber \\
&&\sum_{n=1}^\infty \frac{(-1)^{n+1}}{n^2}
\exp \left( - \pi n\left|\frac{m^2c^3}{\hbar e E}\right|\right).
\label{bpp2}
\end{eqnarray}
A uniform electric field can thereby excite the charged boson oscillators emitting pairs \begin{math}
(\pi^+ \pi^-) \end{math} from the vacuum. The electric field does the work required to {\em break down the
vacuum}.

\subsection{Charged Particle Paths \label{cpp}}

Let us here consider how the zero-point  energy is expressed in terms of paths forward in time (particle)
and backward in time (anti-particle).
Let us at first work in one space and one time (1+1) dimensions. With a
small modification, this leads to a correct description in physical three space and one time (3+1)
dimensions.

In (1+1) dimensions, the energy-momentum relation reads
\begin{equation}
{\cal E}^2-c^2p^2=(mc^2)^2 .
\label{cpp1}
\end{equation}
Since energy is force times distance and momentum is force times time, Eq.(\ref{cpp1}) reads
\begin{equation}
(eEx)^2-c^2(eEt)^2=(mc^2)^2 ,
\label{cpp2}
\end{equation}
or in terms of the particle acceleration \begin{math} a \end{math},
\begin{equation}
a= \frac{eE}{m},
\label{cpp3}
\end{equation}
Eq.(\ref{cpp2}) reads
\begin{equation}
x^2-c^2t^2= \left( \frac{c^2}{a} \right)^2
\label{cpp4}
\end{equation}
that describes classical paths. The particle path forward in time is
\begin{equation}
x_+(t)=\sqrt{c^2t^2+(c^2/a)^2}
\label{cpp5}
\end{equation}
while the anti-particle path backward in time is
\begin{equation}
x_-(t)=-\sqrt{c^2t^2+(c^2/a)^2}
\label{cpp6}
\end{equation}
Pair production at {\em time zero} requires a space-like transition from \begin{math} x_-(0)=-(c^2/a)
\end{math} to
\begin{math} x_+(0)=(c^2/a) \end{math} along the semicircle in Euclidean time
\begin{math} t_{\cal E} \end{math}, i.e. Eq.(\ref{cpp4}) reads in Euclidean
time
\begin{equation}
x^2+c^2t_{\cal E}^2=\left( \frac{c^2}{a} \right)^2 .
\label{cpp7}
\end{equation}
The arc length of the semicircle is \begin{math} s=\pi (c^2/a) \end{math} giving rise to the Euclidean
action
\begin{equation}
W=m\, c \, s=\pi \, \frac{mc^3}{a} =\pi \,\frac{m^2c^3}{eE} .
\label{cpp8}
\end{equation}
The boson weight of such pair production processes summed over the number
\begin{math} k \end{math} of pairs produced is related to the partition
function
\begin{equation}
{\cal Z}=\sum_{k=0}^\infty (-1)^k e^{-kW/\hbar } =  \frac{1}{1+e^{-W/\hbar }} .
\label{cpp9}
\end{equation}
The factor of \begin{math} -1 \end{math} for each semicircle means a Bose factor of one for each circle.
Since the rate of change of momentum is equal to the force, \begin{math} dp/dt = eE  \end{math}, the
transition rate per unit time per unit length \begin{math} \Gamma_1 \end{math} is given by
\begin{eqnarray}
\Gamma_1 dt = \frac{dp}{2\pi \hbar } \, (-\ln {\cal Z}) ,
\end{eqnarray}
i.e.
\begin{eqnarray}
\Gamma_1 &=& \frac{eE}{2\pi \hbar } \,
\ln \left[1+e^{-(\pi m^2c^3/\hbar eE)}\right]
\nonumber \\
&=& \frac{eE}{2\pi \hbar } \,
\sum_{n=1}^\infty \frac{(-1)^{n+1}}{n}\ e^{-n(\pi m^2c^3/\hbar eE)}.
\label{cpp10}
\end{eqnarray}

By taking the momentum perpendicular to the electric field into account, the (3+1) dimensional result
follows from Eq.(\ref{cpp10}),
\begin{eqnarray}
{}\hspace{-7mm}\Gamma &=& \frac{eE}{2\pi \hbar } \, \int
\frac{d^2 {\bf p}_\perp}{(2\pi \hbar)^2} \times 
\nonumber\\
&&\sum_{n=1}^\infty \frac{(-1)^{n+1}}{n}
\exp \left( - \pi n\left|\frac{m^2c^3+cp_\perp ^2}{\hbar e E}\right|\right) =
\nonumber \\
{}\hspace{-7mm} &{}& \frac{c}{8\pi^3}\left(\frac{eE}{\hbar c}\right)^2
\sum_{n=1}^\infty \frac{(-1)^{n+1}}{n^2}
\exp \left(- \pi n\left|\frac{m^2c^3}{\hbar e E}\right|\right),~~~~
\label{cpp11}
\end{eqnarray}
in agreement with Eq.(\ref{bpp2}).

It is not difficult to write the transition rate for producing pairs wherein the charged particles have
spin \begin{math} s \end{math}, i.e.   $s=0,\ 1,\ 2,\ 3,\ \cdots $
for bosons,
and  $s=1/2,\ 3/2,\ 5/2,\  \cdots  $
for fermions.
The
statistical index may be defined as
$\eta_s = \exp\big[i\pi (2s+1)\big]$,
$\eta_s = -1 \ \ \   {\rm for~bosons}$,  \,  $\eta_s = +1 \ \ \  {\rm for~fermions}$.
From a QFT viewpoint, the statistical index is related to the commutation
or anti-commutation relation between creation and destruction operators
\begin{equation}
[a,a^\dagger ]_{\eta_s}=a a^\dagger +\eta_s a^\dagger a =1
\label{cpp12c}
\end{equation}
For arbitrary spin, Eq.(\ref{cpp11}) may be argued from the factor
\begin{math} -\eta_s  \end{math} for each closed circle loop to be
\begin{eqnarray}
\Gamma &=& \frac{(2s+1)c}{8\pi ^3}\left(\frac{eE}{\hbar c}\right)^2 \times
\nonumber \\
&&\sum_{n=1}^\infty \frac{\eta_s^{(n+1)}}{n^2}
\exp \left( - \pi n\left|\frac{m^2c^3}{\hbar e E}\right|\right).
\label{cpp12d}
\end{eqnarray}
Eq.(\ref{cpp12d}) has been discussed in the literature \cite{Bialas:1984}.

Finally, the Euclidean action \begin{math} W \end{math} may be associated with an entropy \begin{math} S
\end{math} via
\begin{equation}
\frac{W}{\hbar }=\frac{S}{k_B}=
\pi \, \frac{m^2 c^3}{\hbar eE}.
\label{cpp12}
\end{equation}
The derivative of the entropy with respect to the rest energy determines the reciprocal temperature
\begin{equation}
\frac{1}{c^2}\, \frac{dS}{dm} = \frac{1}{T}
\ \ \ \Rightarrow \ \ \ k_BT =  \frac{\hbar eE}{2\pi mc} .
\label{cpp13}
\end{equation}
In terms of the {\em acceleration} of the charged bosons, there exists an effective temperature
\cite{Unruh:1976},
\begin{equation}
k_BT = \frac{\hbar a}{2\pi c}
\label{cpp14}
\end{equation}
of the environment inducing position fluctuations equivalent to the energy fluctuations in the rest frame
of the applied electric field (the Unruh effect or  Unruh temperature).

Let us close by observing that  the central results of this and the previous Section
are not new. For example, the spin zero charged boson pair production rate in
Eq.(\ref{bpp2}), as well as its generalization to the general spin $s$ charged particle pair production rate
in Eq.(\ref{cpp12d}) are  well known. However, the derivations, physical pictures and consequences 
of zero-point  oscillations are to our knowledge original. The notion of zero-point  energy in relativistic QFT is
made real by the particle and anti-particle content of the theory.

\section{Photon two-slit \label{photo}}

We turn now to the second part of our discussion focusing on the forward \&  backward in time motion
formalism for  two-slit photon phenomena. In particular we generalize the forward \& backward in time
motion formalism developed in \cite{Blasone:1998} for non-relativistic two-slit interference processes to
the relativistic case of two-slit processes for photons.

There are several  motivations for such an extension:

\noindent -- there is a well defined radiation QFT with a classical limit called the Maxwell theory.

\noindent -- there is no mass gap for photons contrary to the electrons. A photon is its own anti-particle. Thus,
both forward and backward motions must be there anyway, as  discussed in  Sec.~\ref{Maxwell}.

\noindent -- experimentally, both classical and quantum optics are amongst the most studied subjects in physics;
 and not only theoretically.

Let us first consider a massless spin zero (scalar wave). If Planck, Einstein and Bose could invoke it for
the black-body radiation for example, and then, after computing the radiation energy density, they
multiplied their results by 2 to take care of the two polarizations, we are in good company. However, as we
shall see later, spin is neither harmless nor a trivial complication.

\subsection{Spin zero, massless radiation \label{spinless}}

Under the assumptions of our earlier paper \cite{Blasone:1998}, the diffraction limit formula (Eq.(27) of
ref.\cite{Blasone:1998}) would still read,  {\it mutatis mutandis}, for photons (see Appendix \ref{app2}):
\begin{equation}
\label{1}
P_\gamma(x; D) \approx\ \frac{4}{\pi \beta K_\gamma x^2} \, \cos^2(K_\gamma x) \, \sin^2(\beta K_\gamma x),
\end{equation}
with the following replacement for the definition of $K_\gamma$ compared to $K_{electron}$ (a particle of
mass $M$)
\begin{eqnarray}
\label{2}
K_{electron} = \frac{M v d}{\hbar D}\ \Longrightarrow\\ K_\gamma = \frac{p d}{\hbar D} =
\frac{2\pi}{\lambda} \frac{d}{D} , \label{2a}
\end{eqnarray}
where $p = 2\pi \hbar/\lambda$ is the (mean)-momentum and $\lambda$ the wavelength of the photon.
Also, $w$ is the size of the slit, $2d$ is the distance between the two slits, $\beta = w/d$ and $D$ is the distance of the
screen  from the source.
Thus, the diffraction pattern remains exactly as before.

It is worthy of note that in the extreme limit  $\beta d/\lambda  = w/\lambda\ \to 0$, the quantity $P(x,
D)/\beta K_\gamma$ has a finite limit:
\begin{equation}
\label{2.1}
\left[ \frac{P(x, D)}{\beta K_\gamma} \right]_{\beta=o} = \frac{4}{\pi} \, cos^2(K_\gamma x),
\end{equation}
there remains {\it just} the expected Young's interference pattern  showing maxima at $x_{max}$ (constructive
interference) and minima at $x_{min}$ (destructive interference) according to the path difference ({\it cf.} Eq. (\ref{2a}))
\begin{eqnarray}
\label{2.2}
&&\hspace{-3mm}\sqrt{D^2 + (x + d)^2} - \sqrt{D^2 + (x - d)^2}\approx\ 2 x \frac{d}{D} \ ,
\end{eqnarray}
so that (with integer $ n = 0, 1, 2,\dots$)
\begin{eqnarray}
 &&\hspace{-3mm} 2 x \frac{d}{D} = \lambda n ~\Longrightarrow\ x_{max} = \frac{\lambda}{2} \frac{D}{d} n \ ,
 \\
&&\hspace{-3mm} 2 x \frac{d}{D} = \lambda \left( n +  \frac{1}{2} \right) 
\Longrightarrow x_{min} =  \frac{\lambda}{2} \frac{D}{d} \left( n +  \frac{1}{2} \right).
\end{eqnarray}

Let us pause here and note that we have derived - in general - the diffraction pattern for a scalar photon,
under the hypothesis of symmetric forward and backward time motion. There is no visible trace of quantum
mechanics left, i.e., there are no factors of $\hbar$ in the intensity distribution $P(x, D)$, even though
we have computed it through a probability amplitude involving an action that is scaled by $\hbar$. This has
to do with the peculiarities of a massless field. Explicitly, for a non-relativistic electron without spin,
$K_{electron} = M v d/(\hbar D)$ in Eq.(\ref{2}) contains $\hbar$, whereas for a massless (scalar) photon,
in $K_\gamma = 2\pi d/(\lambda D)$ in Eq.(\ref{2a}), there is no $\hbar$. For a free massless spin-s field $\phi$
the Hamiltonian reads $H = -i \hbar c {\bf \nabla \cdot S}$ and in the corresponding Schr$\ddot o$dinger
equation
\begin{equation}
\label{2M}
i \hbar \frac{\partial \phi}{\partial t} = H \phi,
\end{equation}
$\hbar$ drops out. By contrast, even for a free Dirac field $\psi$ of mass $M$ for instance,
\begin{equation}
\label{2N}
i \hbar \frac{\partial \psi}{\partial t} = [(-i\hbar c) \gamma_5 ({\bf \Sigma \cdot \nabla }) + \beta \, M
c^2 ]\psi,
\end{equation}
$\hbar$ does {\it not} factor out: for the simple reason, that mass destroys scale invariance 
(of course, in Eq. (\ref{2N}) $\beta = \gamma_4$  and $\Sigma_i = -i \gamma_{j}\gamma_{k}$, 
with cyclic $i,j,k = 1,2,3$).

To recapitulate, we expect - at least formally - that there is no distinction between a free classical
massless field and its counterpart quantum field. The intensity distribution verifies it exactly for the
time-symmetric propagation of a scalar massless field. Quantum mechanics tells us, in Feynman's language,
that if we had such a {\it photon gun} firing at the two slits one photon at a time, we should find the
diffraction pattern in the observed intensity when both slits are open as obtained in Eq.(\ref{1}). And
this expression has no $\hbar$ in it.


Of course, once non-commuting spins are introduced, say for a massless photon of spin 1, the quantum and
classical theories need not be equivalent. For example, the spin would precess for a given mean momentum
$p$ at an angular frequency $\Omega = pc/\hbar = 2\pi c/\lambda$ in a plane perpendicular to the direction
of motion \cite{Siva1:2004}. We consider this in the following Section.

\subsection{Spin-one radiation \& Maxwell theory \label{spin1}}

Much of the formalism discussed below can be found in \cite{Siva1:2004}. While the emphasis in
\cite{Siva1:2004} was upon the non-commutative photon coordinates and on methods for its revelation through
two photon processes, here we shall be focussing on the behavior of a single photon for two-slit
arrangements to shed light on forward and backward in time photon propagation.

There are two sets of (non commuting) coordinates and  velocity operators ${\bf V}$ that are given by
\begin{equation}
\label{11}
{\dot{\bf X}} = {\bf V} = \frac{\partial \mathcal{H}}{\partial {\bf p}} = c \beta \ {\bf S}.
\end{equation}
Due to the divergence condition Eq.(\ref{3}), the motion of the field $\Psi$ is confined to the plane
perpendicular to the momentum. Thus, only the motion of coordinates and velocities  in the plane
perpendicular to ${\bf p}$ are of physical relevance here. For example, if the momentum is directed along
the z-axis, Eq.(\ref{11}) tells us that the x- and y-components of the velocities do not commute
\begin{equation}
\label{12}
[V_{+, 1} , V_{+,2} ] = i c^2 \Lambda = [V_{-, 1} , V_{-,2} ],
\end{equation}
where $\Lambda = \pm 1$ is the helicity of the photon. The mixed commutator
\begin{equation}
\label{13}
[V_{+, 1} , V_{-,2} ] = i c^2 \beta \Lambda,
\end{equation}
shall not be discussed here, as it does not enter the discussions to follow.

The non-commutativity of the photon position coordinates lying in a plane orthogonal to the direction of
its motion Eq.(\ref{11}) has been throughly discussed in Refs.\cite{Siva1:2004,Siva2:2004}. In particular,
for a photon of helicity +1 moving along the z-axis, the $X_1,X_2$ coordinates of the photon precess about the
z-axis with a frequency $pc/\hbar$ and the radius ${\bf R}^2 = (X_1^2 + X_2^2)$ is quantized:
\begin{equation}
\label{14}
R_n^2 = \left (\frac{\hbar}{p} \right)^2 (2n +1) = \left (\frac{\lambda}{2\pi} \right)^2 (2n + 1);\
n=0,1,\dots
\end{equation}
Of course, the {\it center} is unspecified, that is why {\it two, parallel, same helicity,} photons were
needed in \cite{Siva1:2004} to allow for a measurement of the quantization in the difference between the
(transversal) positions of the two photons. Incidentally, Maxwell was well aware of two opposite screw
motions (corresponding to the two helicities) about the axis of propagation. He was only missing the names
{\it photon} and {\it spin} (and possible quantization conditions) for the EM waves \cite{Maxwell-I}.

Returning to our two-slit arrangement for a single photon, we can try to obtain some information from the
above quantization condition. As in the constant magnetic field Landau level problem, there is a huge
degeneracy introduced by the uncertainty in the center of the coordinates. The number of states/area for
the magnetic case is well known to be $eB/(2\pi \hbar c)$.
The density of transversal states is given by
\begin{equation}
\label{15}
\rho = \frac{1}{2\pi} { \left( \frac{2\pi}{\lambda} \right)^2}  = \frac{2\pi}{\lambda^2}.
\end{equation}
In our problem, each of the slits of total width $w$ can be considered as a circle (in the $x,y$ plane) of
area $A = \pi (w/2)^2$. Hence, we can estimate (semi-classically) the total number traversing each slit to
be
\begin{equation}
\label{16}
N = \rho A = \frac{\pi^2}{2} \left ( \frac{w}{\lambda} \right)^2.
\end{equation}
Thus, at least theoretically it would appear as if we can confine the transversal photon coordinates to be
in its {\it ground} state $n = 0$ for $w < (\lambda/\pi)$.\\ There is a matter of principle involved here.
For a scalar wave diffraction pattern, Eqs.(\ref{1}),(\ref{2}), there appears to be no theoretical lower
limit to $w$ apart from $w\ll d\ll  D$. On the other hand, for a spin-one photon, there is a quantum
constraint. Can it be measured?

With micro/nano technology, both the fabrication of apertures
small enough as well as procurement of polarized light of wavelengths smaller than the size of the
apertures should be possible.
With such  setups, the very
interesting
fine structure in the diffraction pattern can be investigated as the width $w$ is lowered for a
fixed wavelength $\lambda$.  It seems to us, on the basis of the discussion in this
paper,  that an experiment in such a setup might be very worthwhile.

What one can find in classic texts such as \cite{Born:1970} are diffraction patterns in the limit where
$\lambda \ll w \ll d \ll D$. What is interesting, and to us
at least intriguing, is that even for mercury
light of wavelength $\lambda \sim 5.79 \times 10^{-5}\ cm$ passing through a single aperture $w \sim 0.6\
cm$ , four or five diffraction minima are clearly visible. Thus, in the diffractive part of the spectrum
\begin{equation}
\label{17}
S(x) = \left( \frac{\sin \eta}{\eta} \right)^2;\ \ \eta = \frac{2\pi x w}{\lambda D}.
\end{equation}
At the first minimum say, $\eta_1 = \pi$. Translated into the vertical distance $x$ on the screen to the
distance of the screen $D$ from the source, one finds $x_1 \sim 10^{-4} D$. Thus, even a meter away, the
value of $x_1 \sim 10^{-2}\ cm$. To  us it is remarkable that it can be measured so well.

In any event, if small apertures of size $w \sim (10^{-5}\div 10^{-4})\ cm$, can be fabricated, then $x_1
\sim D$ and measurements might be easier.  A discussion on experimental issues is undertaken in the next
Section.

\subsection{Some experimental issues \label{past}}

In this subsection we shall discuss a few interference and diffraction experiments done in the past by way of
comparison to the proposed two slit experiments for photons in the present paper.

{\em Magnetic fields and the Quantum Hall Effect:} Let us begin by recalling the well-known fact that once
a magnetic field is introduced via a vector potential, the components of velocity ${\bf v} = ({\bf p} - e
{\bf A}/c)/M$ - even for a non-relativistic electron - do not commute
\begin{eqnarray}
\label{m1}
[v_i, v_j] = \frac{ie\hbar}{M^2 c} \epsilon_{ijk} B_k = - \frac{e\hbar}{M^2 c} ({\bf S\cdot B}).
\end{eqnarray}
In particular, for ${\bf B} = B\ \hat{k}$, we have
\begin{eqnarray}
\left[ \frac{v_1}{c} ,\frac{v_2}{c} \right] =
 - \frac{\hbar \omega_B}{M c^2}  \equiv - \Delta,
\end{eqnarray}
with $\omega_B ={eB}/(Mc)$.
If one compares Eq.(\ref{m1}) for the non-commuting components of the electron velocity (that are
perpendicular to the magnetic field),
with the corresponding non-commuting
components of the photon velocity (that are perpendicular to the direction of motion of the photon) given
in Eq.(\ref{12}), one finds the right hand side of the commutator for the electron a rather small value
$\Delta \ll 1$, whereas for the photon the factor is unity. It is for this reason that (for the case of an
electron) one needs high magnetic fields and low temperatures so that thermal fluctuations do not wash out
the quantum effects for an electron. For example, integer quantum Hall steps were made visible
experimentally \cite{Klitzing:1980}, with $B = 18\ Tesla$ and at a low temperature $T = 1.5\ K$.(The value
of $\Delta \sim 10^{-9}$ for this experiment).\\ To observe fractional quantum Hall steps \cite{Tsui:1982},
even higher fields ($B \sim 35\ Tesla$) and milli-Kelvin temperatures were necessary. For a detailed
derivation of quantum Hall steps in the context of QED, see \cite{Friedman:1984} and for a review see
\cite{Widom:1987}.\\ The relevant point of the above discussion for the present paper is that for the
photon case there are no small factors such as $\Delta$, and thus visibility of the proposed quantum
effects for the photon are not afflicted by background thermal fluctuations and experiments at room
temperatures should be adequate.

{\em Cold neutron experiments:} Several very cold neutron diffraction experiments have been performed and
they have been excellently reviewed in \cite{Zeilinger:1988}. The wavelength of the neutrons in such
experiments is typically  $\lambda_{neutron} \approx\ 20 \mathring{A}$, to be compared with $3900
\mathring{A} \leq \lambda_{visible-light} \leq 7000 \mathring{A}$. The relevant slit widths in such
experiments were about $w = 20\ \mu m = 10^4\ \lambda_{neutron}$. In the experimentally covered regime $w
\gg \lambda$, the observed diffraction patterns are in good agreement with their theoretical expectations.
The prospect of future cold neutron experiments in the opposite regime $w \ll \lambda$ is rather remote. On
the other hand, as we outline below, for photons in the visible spectrum with wavelengths over two hundred
times larger than the cold neutron wavelengths, fabrication of needed slit widths of sufficiently small
size (say $\leq 0.1\ \mu m$) may not be technically so daunting.

{\em Photon double slit experiments:} As we have discussed at length in Sec.\ref{spin1}, there is a
fundamental difference between the propagation of a massless scalar (spin zero) wave through two slits as
compared to that of a massless vector (spin one) wave. This should not be surprising as the former has ``no
directional pointers'', whereas the latter does have one through the direction of the spin. In practice,
there is a precession of the spin at a frequency $\omega = pc/\hbar$ in a plane perpendicular to the
direction of motion of the photon. Since the transversal velocities of the photon do not commute ({\it
cf.} Eq.(\ref{12})), the transversal positions of the photon satisfy a quantum Pythagoras theorem ({\it
cf.} Eq.(\ref{14})) with an arbitrary center of the circular orbits.\\ Thus, in contrast to a scalar wave,
there is degeneracy constraint for a physical EM vector-wave traversing a slit of width $w$. The number of
states is given by Eq.(\ref{16}) to be $N = ({\pi^2}/{2}) ({w}/{\lambda})^2$. Thus, for small
enough slits $w \leq \lambda/\pi$, we can limit the transversal quantum number $n$ to the ground state
($n=0$).\\ To be concrete, let us consider mercury yellow light of wavelength $\lambda = 0.58\ \mu m$. The
standard diffraction pattern as expected have been confirmed for ``large'' slits of width $w \sim 0.6\ cm$
\cite{Born:1970}. Our proposal is to vary the slit width $w$ and observe the change in the diffraction
pattern specially once it is reduced to $0.1\ \mu m$ or even lower. (Of course, respecting $w\ll d$, the
distance between the two slits). Such a region to our knowledge has not been explored previously and that
is our suggestion.

\section{Conclusions \label{Con}}

In the present paper, as in our earlier papers on the subject, we have shown that both forward and
backward motions in time are essential for a proper description of a particle's motion from its classical
to its quantum counter part. For the important case of a photon, it plays a particularly decisive role. A
free photon described by the Maxwell equations in its inherent time symmetric aspect has been shown to be
essential for obtaining the correct Planck thermal radiation distribution with the zero-point energy. As
stressed in the body of the text, as a photon is its own anti-particle, its motion in forward and backward
motions in time 
is often overlooked. Thus, various aspects of the dynamics of a charged particle for which
forward and backward in time motions are distinct have been considered in detail. \\ When applied to the
case of a photon, considered first as a scalar
field,  standard expressions for interference and diffraction have been obtained. On the other hand, when
extended to the realistic case of a spin 1 photon, the non-commutativity of the spin components, induce
non-commutativity in the components of the photon position coordinates. As the commutator between two such
coordinates is proportional to the square of the wave-length, the intrinsic uncertainty in the position of
a photon is proportional to its wavelength. We have shown here that it can manifest itself through changes
in the interference pattern of a two-slit photon experiment as the width of a slit is lowered below the
wavelength of the photon.\\ Our formalism also provides an understanding of why a strict localization of a
photon - to better than its wave length - runs into serious difficulties \cite{Mandel:1995}. The
experiments suggested in the present paper should provide definite light not only on the validity of the
formalism but also on the  fundamental subject of photon localizability.

\section*{ACKNOWLEDGEMENTS}
YS would like to thank the Department of Physics and Geology at the University of Perugia for its
hospitality.

\appendix


\section{Gamma Function Regulation \label{gfr}}

The Gamma function is defined in the \begin{math} {\Re e}\{z\}> 0 \end{math} part of the complex plane as
\begin{equation}
\Gamma (z)=\int_0^\infty t^z e^{-t} \left(\frac{dt}{t}\right) ,
\label{gfr1}
\end{equation}
from which we find for \begin{math} a>0 \end{math} the identity
\begin{equation}
a^{-z}=
\frac{1}{\Gamma(z)}\int_0^\infty s^z e^{-as} \left(\frac{ds}{s}\right).
\label{gfr2}
\end{equation}
In other regimes, \begin{math} \Gamma (z) \end{math} is defined by analytic continuation. This analysis is
often assisted by the identity
\begin{equation}
\Gamma (1+z)=z\Gamma (z).
\label{gfr3}
\end{equation}
For example, by putting \begin{math} z=-(1/2) \end{math} in Eq.(\ref{gfr3}), one finds
\begin{equation}
\Gamma \left(\frac{1}{2}\right)=\sqrt{\pi } \ \ \ \ \ \Rightarrow
\ \ \ \ \ \Gamma \left(-\frac{1}{2}\right)=-2\sqrt{\pi}\ .
\label{gfr4}
\end{equation}
Eqs.(\ref{gfr2}) and (\ref{gfr4}) lead to a {\em formally divergent integral}
\begin{equation}
\sqrt{a}=- \frac{1}{2\sqrt{\pi }}
\int_0^\infty e^{-as} \left(\frac{ds}{s^{3/2}}\right).
\label{gfr5}
\end{equation}
For those readers who find it strange in Eq.(\ref{gfr5}) to put a finite positive quantity equal to a
negative infinite quantity, we invite the reader to prove the following \\ {\bf Theorem:} For any
\begin{math} a>0 \end{math} and
\begin{math} b>0 \end{math}
\begin{equation}
\sqrt{a}-\sqrt{b}= \frac{1}{2\sqrt{\pi }}  \int_0^\infty
\left[e^{-bs}-e^{-as}\right]\left(\frac{ds}{s^{3/2}}\right).
\label{gfr5c}
\end{equation}
Subtractions will be made below. Eqs.(\ref{pion3}) and (\ref{gfr5}) yield
\begin{eqnarray}
\hspace{-7mm}U_0(B)&=&- \frac{eB}{8\pi^{5/2}}
\sum_{n=0}^\infty \int_{-\infty}^\infty
dk \int_0^\infty \left(\frac{ds}{s^{3/2}}\right) \times
\nonumber \\
&&\exp\left[-\kappa ^2 s -k^2 s -(2n+1)\left|\frac{eB}{\hbar c}\right|s\right]
\nonumber \\
&=&- \frac{\hbar c}{16\pi^2} \int_0^\infty
\frac{ds}{s^3}e^{-\kappa ^2s}
\frac{(eBs/\hbar c)}{\sinh(eBs/\hbar c)}
\label{gfr6}
\end{eqnarray}
that is still divergent. One then subtracts the vacuum zero-point  oscillations when the magnetic field is
zero,
\begin{equation}
\tilde{U}(B)=U_0(B) - U_0(0).
\label{gfr7b}
\end{equation}
The zero-point  oscillation energy per unit volume due to vacuum particle anti-particle pairs, say
\begin{math} (\pi^+ \pi^-) \end{math}, virtual magnetic moments are thereby
\begin{eqnarray}
\tilde{U}(B)= \frac{\hbar c}{16\pi^2} 
\int_0^\infty \frac{ds}{s^3}\ e^{-\kappa ^2s}
\left[1- \frac{eBs/\hbar c}{\sinh(eBs/\hbar c)}\right],
\label{gfr8}
\end{eqnarray}
which still is divergent but only as a logarithm at small distance squared as \begin{math} s\to 0^+
\end{math}. Once the divergences are only logarithmic, one may pass from {\em regularization} to {\em
renormalization}.

For what concerns the charge {\em renormalization} procedure we observe that in quantum electrodynamics one
starts with charges and fields described for the problem at hand by \begin{math} e_0  \end{math} and
\begin{math} B_0 \end{math}. Both the fields and charges have to be
renormalized by considering the vacuum polarization contributions \cite{Schwinger:1951,Umezawa:1951} in such a way that \begin{math} e_0 B_0=eB \end{math} and thus divergent logarithms are
buried. The vacuum energy density for scalar boson fields is reported in Sec.~\ref{pion}.

\section{Derivation of Eq.(\ref{1}) \label{app2}}

Here we exhibit the simple calculations through which Eq.(\ref{1}) is obtained for a scalar (spin 0)
massless photon. It fixes the precise assumptions and the notation. Using the formalism and notation of
\cite{Blasone:1998}, we have Eq.(18) of \cite{Blasone:1998},  but with $H = (1/2) Mv^2$ for a non-relativistic particle replaced
by $H_\gamma = E_\gamma = pc = 2 \pi c /\lambda$. So that, the classical action for a scalar photon reads
$A_\gamma = p\, c \, t$. Thus, Eq.(19) of \cite{Blasone:1998}, for the amplitude is replaced by
\begin{equation}
\label{A1}
P_\gamma(x,t) = \frac{1}{\lambda x} \, \int_{-\infty}^{+\infty}\!\! \! dx_+  \int_{-\infty}^{+\infty}\!\!\! dx_{-} \
e^{i\mathcal{S}_\gamma} (x_+ | \rho_o| x_-),
\end{equation}
where
\begin{eqnarray}
\label{A2}
&&\mathcal{S}_\gamma = \mathcal{S}_+ - \mathcal{S}_-;\\ &&\mathcal{S}_+ = H_{(+,\gamma)} (t_+/\hbar) =
\frac{pc}{\hbar} \, |{\bf x} - {\bf x}_+ |/c;\\ &&\mathcal{S}_- = H_{(-,\gamma)} (t_-/\hbar) =
\frac{pc}{\hbar} \, |{\bf x} - {\bf x}_- |/c.\\ &&\mathcal{S} = \frac{2\pi}{\lambda}  \,  \Big[|{\bf x} - {\bf
x}_+ | - |{\bf x} - {\bf x}_- |\Big].
\end{eqnarray}
Using $|{\bf x} - {\bf x}_i| = \sqrt{D^2 + (x - x_i)^2}$, in the diffraction limit of large $D$ (the
Fraunhoffer limit valid up to quadratic order \cite{Born:1970}), the effective action reduces to
\begin{equation}
\label{A3}
\mathcal{S}_\gamma \approx\ - \frac{2\pi x}{\lambda D} \, (x_+  - x_-).
\end{equation}

The choice and construction of the initial density matrix for the two-slit arrangement with each slit of
size $w$ placed a distance $2d$ apart, proceeds identically as in [Eqs.(15), (22), (23), (24) of \cite{Blasone:1998}]:
\begin{eqnarray}
\label{A4}
{}\hspace{-3mm}&&(x_+ | \rho_o| x_-) = \psi_o^*(x_-) \psi_o(x_+);\nonumber\\
&&\psi_o(x) = \frac{1}{\sqrt{2}} \Big[ \phi(x- d) + \phi(x + d)\Big];\nonumber\\
&&\phi(x) =  \frac{1}{\sqrt{w}};\ \ |x| < \frac{w}{2};\ {\rm and}\  = 0 ~\ {\rm otherwise},
\end{eqnarray}
so that as in our original paper \cite{Blasone:1998}
\begin{eqnarray}
\label{A5}
\hspace{-2mm}(x_+ | \rho_o| x_-) &=& \frac{1}{2} \Big[ \phi(x_+ - d) \phi(x_- - d)  \nonumber\\
&+& \, \phi(x_+ + d) \phi(x_- + d)
 + \phi(x_+ - d) \phi(x_- + d)\nonumber\\
 & +&  \phi(x_+ + d) \phi(x_- - d) \Big].
\end{eqnarray}

Using Eqs.~(\ref{A3}, \ref{A4}, \ref{A5}), the integrals involved in Eq.(\ref{A1}) for the computation of
$P_\gamma(x, D)$ are finite-range Fourier transforms that lead to a product of (Young's interference
factor) $Y(Kx)$ with (Fraunhoffer diffraction factor) $F(\beta K x)$, where $K = ({2\pi}/{\lambda})({d}/{D})$ and $\beta = w/d$: %
\begin{eqnarray}
\label{A6}
P_\gamma(x, D) &=& \frac{4 \beta K}{\pi} Y(Kx) F(\beta K x);\\ Y(K x) &=& \cos^2(Kx); \\ F(\zeta) &=& \left[ \frac{\sin \zeta}{\zeta} \right]^2;\  \zeta =
\beta K x ;
\end{eqnarray}
the result quoted in Eq.(\ref{1}).

The dimensionless quantity $P_\gamma/(\beta K)$
\begin{equation}
\label{A7}
\frac{P_\gamma(x, D)}{\beta K} = \frac{4}{\pi} \, Y(Kx) F(\beta K x),
\end{equation}
has a smooth limit as the size of each slit $w \to 0$. In this limit, the diffraction pattern disappears,
leaving behind just the Young's interference pattern. Of course, for visible light this is only a deceptive
limit unless -  as discussed in the text - apertures can be constructed for which $w \ll  \lambda$, an
arduous task.

We mention in passing that if one assumes two circular apertures each of (radius $w/2$), $P(x,D)$ can be
computed exactly as above: the only change is in the Fraunhoffer diffraction function that now reads
\begin{equation}
\label{A8}
F_{circle} =\left [ \frac{2 J_1(\eta)}{\eta} \right]^2;\ ~~ \eta =  \frac{2\pi}{\lambda} \, \frac{w}{D} \,
R.
\end{equation}
In this case, we expect circular (bright and dark) rings, with a  maximum at $R = 0$, and minima at various
$R$, corresponding to the zeroes of the Bessel function of the first kind $J_1(\eta)$; see
\cite{Born:1970}. 

\end{document}